# Ferromagnetic InMnAs on InAs Prepared by Ion Implantation and Pulsed Laser Annealing

Shengqiang Zhou[1], Yutian Wang[1,3], Zenan Jiang[1], Eugen Weschke[2], and Manfred Helm[1,3]

[1] *Institute of Ion Beam Physics and Materials Research, Helmholtz-Zentrum Dresden-Rossendorf, P. O. Box 510119, Dresden 01314, Germany*

[2] *Helmholtz-Zentrum Berlin für Materialien und Energie, Wilhelm-Conrad-Röntgen-Campus BESSY II, Albert-Einstein-Straße 15, D-12489 Berlin, Germany*

[3] *Technische Universität Dresden, 01062 Dresden, Germany*

Abstract:

Ferromagnetic InMnAs has been prepared by Mn ion implantation and pulsed laser annealing. The InMnAs layer reveals a saturated magnetization of 2.6 $\mu_B$/Mn at 5 K and a perpendicular magnetic anisotropy. The Curie temperature is determined to be 46 K, which is higher than those in previous reports with similar Mn concentrations. Ferromagnetism is further evidenced by the large magnetic circular dichroism.

Diluted magnetic semiconductors (DMS) have been under intensive investigation during the last decade [1]. Among them, GaMnAs has been mostly studied and has become a kind of model system. Compared with GaAs, the semiconductor InAs has a much smaller bandgap (0.4 eV) [2] and a much larger hole mobility [3, 4]. Although two decades have passed since the first growth of ferromagnetic InMnAs (in fact, the first III-V diluted ferromagnetic semiconductor) by Munekata *et al.* [5, 6], its preparation by molecular beam epitaxy (MBE) remains challenging. In the early work by Munekata *et al.*, epitaxial InMnAs films grown on InAs buffered GaAs substrates are either paramagnetic or with a Curie temperature ($T_C$) below 10 K [5, 6]. The reason is the large lattice mismatch between InMnAs and GaAs, resulting in a large amount of n-type defects. By choosing proper buffer layers (GaAlSb or AlAsSb) [3, 4, 7] to reduce the lattice mismatch or by reducing the thickness of the films [8], ferromagnetic InMnAs films with higher $T_C$ have been obtained. The highest $T_C$ of 90 K was reached when the Mn concentration was 10% and combined with low-temperature annealing [7]. Alternatively, InMnAs has been grown homoepitaxially on InAs by metal organic vapor phase epitaxy (MOVPE). Those InMnAs films show a hole-mediated ferromagnetic phase with $T_C$ < 30 K, but a much smaller spin polarization than GaMnAs [9].

In this Letter, we show the preparation of ferromagnetic InMnAs on InAs by Mn ion implantation and pulsed laser annealing, previously proposed by Scarpulla *et al.* [10, 11]. The advantage of pulsed laser annealing is the high process temperature within nano-second, eliminating n-type defects [10]. The obtained InMnAs films show distinct perpendicular magnetic anisotropy and a Curie temperature of 46 K. We estimate the X-ray magnetic circular dichroism to be as large as 51% at the Mn $L_3$ edge.

Intrinsic InAs(001) wafers were implanted with 100 keV Mn ions at 240 K by liquid nitrogen cooling. The Mn fluence was $8\times10^{15}$/cm$^2$, resulting in a peak concentration of 5% (or an average concentration of 4% over a depth of around 80 nm). The samples were annealed with a single pulse and energy density of 0.15–0.25 J/cm$^2$ from a XeCl excimer laser (Coherent COMPexPRO201, λ=308 nm, 30 ns FWHM). A homogenized 5×5 mm² laser beam was realized by a VarioLas optical system. X-ray absorption spectroscopy (XAS) measurements were performed at the beamline UE46/PGM-1 at BESSY II (Helmholtz-Zentrum Berlin). A magnetic field up to 6 T was applied parallel to the photon helicity and perpendicular to the surface. Magnetic properties were analyzed using a superconducting quantum interference device (SQUID) magnetometer (Quantum Design MPMS).

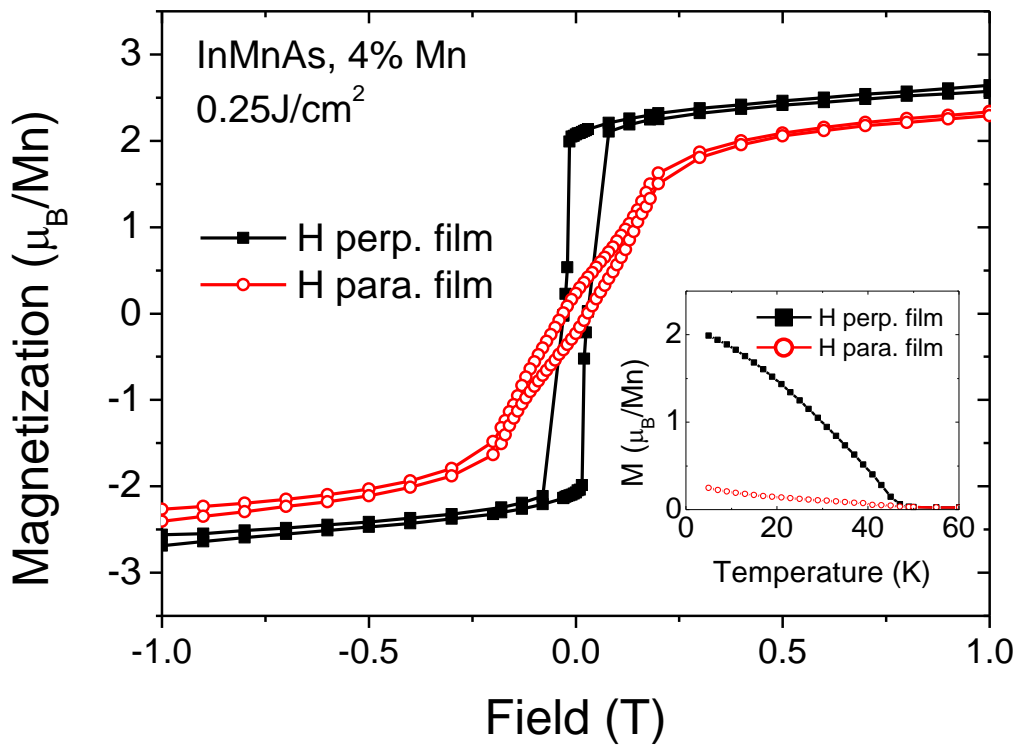

Fig. 1 Field dependence of magnetization measured by SQUID magnetometry. The inset shows the temperature dependence of magnetization measured under 0.005 T but after saturation at 1 T. The field is applied either perpendicular or parallel to the film. The magnetic anisotropy can be clearly observed and the surface normal is the easy axis.

Figure 1 shows the magnetization versus applied field measured at 5 K with the field perpendicular and parallel to the film plane. When the field is perpendicular to the film, a nearly square hysteresis loop is observed. The saturation magnetization is ~ 2.6 $\mu_B$/Mn by considering all implanted Mn ions. This value can be largely underestimated due to two facts. One is the sputtering effect in III-V semiconductors during ion implantation, which will reduce the effective implantation fluence [12]. The second is the surface segregation during pulsed laser annealing [13], which results in a magnetic inert surface layer with a higher Mn concentration. Nevertheless, the estimated magnetization of 2.6 $\mu_B$/Mn is larger than that of InMnAs with $T_C$ of 75 K grown by MBE [7] and it is close to the magnetization of the state-of-the-art GaMnAs [14]. The large saturation magnetization indicates a large activation (substitution) fraction of implanted Mn ions. The inset of Fig. 1 shows the magnetization versus temperature, which was measured while increasing the temperature under a field of 0.005 T, but after saturation by applying 1 T. The Curie temperature ($T_C$) for this sample is around 46 K, which is remarkably high by considering the relatively low Mn concentration of 4 -5%. In Table I, we compare the magnetic properties of our InMnAs films with the state-of-the-art GaMnAs as well as InMnAs and InMnSb prepared by MBE.

Moreover, a clear magnetic anisotropy is observed as shown in Fig. 1 from both field- and temperature-dependent magnetization. The in-plane direction is the magnetic hard axis. The shape of the hard-axis hysteresis loop is very similar to that for GaMnAs. This can be expected since the InMnAs layer is under tensile strain, *i.e.*, with an expanded lattice constant in the in-plane, while a compressed one in the out-of-plane direction [5]. The observed magnetic anisotropy is consistent with carrier-mediated ferromagnetism and can be well understood in the frame of the mean-field model of ferromagnetism mediated by delocalized or weakly localized holes [18, 19].

Table I Comparison between our InMnAs film with GaMnAs, InMnAs, and InMnSb reported by other groups.

| Materials | Mn concentration | Saturation Magnetization | Curie temperature $T_C$ |
|---|---|---|---|
| GaMnAs [14] | 5-9% | 2.6 - 3.4 $\mu_B$/Mn | 90-170 K |
| InMnAs/AlAsSb [7] | 10% | 1.9 $\mu_B$/Mn at 4 K | 75 K |
| InMnAs/InAlAs [15] | 12% | 0.85 $\mu_B$/Mn at 5 K | 45 K |
| InMnSb/InSb [16] | 2.8% | 1.6 $\mu_B$/Mn at 2 K | 8.5 K |
| InMnAs/InAs (this work) | 4-5% | 2.6 $\mu_B$/Mn at 5 K | 46 K |

To further prove the intrinsic nature of the ferromagnetism in InMnAs obtained by ion implantation and pulsed laser annealing, we have performed XAS utilizing circularly polarized light. In Mn $L_{3,2}$ XAS, $2p$ core electrons are excited into the unoccupied $3d$ states. This technique directly probes the electronic structure of the polarized Mn $3d$ band [20]. Moreover, quantitative magnetic information, *e.g.*, spin polarization, can be deduced from the x-ray magnetic circular dichroism (XMCD), *i.e.*, the difference between XAS when changing the helicity of the incoming photons.

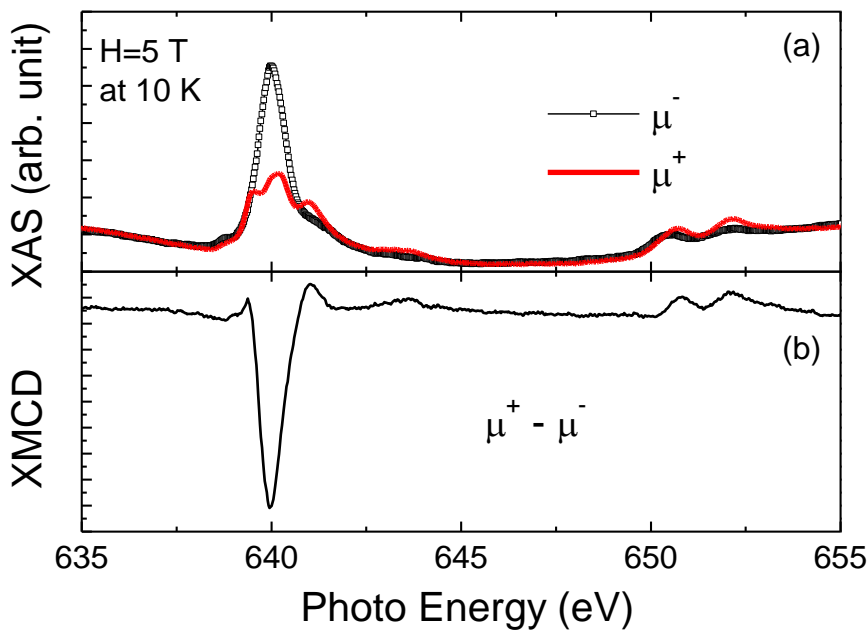

Fig. 2 Mn $L_{3,2}$ total electron yield (TEY) (a) XAS for magnetization and helicity parallel ($\mu^+$) and antiparallel ($\mu^-$) and (b) XMCD ($\mu^+ - \mu^-$) for InMnAs measured at around 10 K under an external field of 5 T applied perpendicular to the surface.

Figure 2(a) presents the Mn $L_{3,2}$ XAS measured in total electron yield (TEY) mode at around 10 K. Right before XAS measurements, the sample was etched in HCl (5%) for 2 min to remove the surface oxide layer [20, 21]. This procedure is crucial to obtain the intrinsic signal from diluted Mn since in the TEY mode one only probes a surface layer of nm thickness. As shown in Fig. 2(a), after etching we obtained very similar spectra as was reported for ferromagnetic GaMnAs [20]. This indicates that the bonding and *p-d* exchange in InMnAs and in GaMnAs are similar. The difference between the two spectra in Fig. 2(a) is larger than 50% at the Mn $L_3$ peak. For InMnAs prepared by MOVPE [9], the difference was only 15% at 5 K and 2 T where the XAS resembles the feature of GaMnAs before removing the surface oxide layer. Figure 2(b) shows the XMCD difference spectrum. The XMCD signal can be quantified using the asymmetry, defined as $(\mu^+ - \mu^-)/(\mu^+ + \mu^-)$. We estimate the maximum asymmetry to be 51% at the $L_3$ peak after taking into account the 90% polarization of the photons, which is as large as that for the state-of-the-art GaMnAs [20].

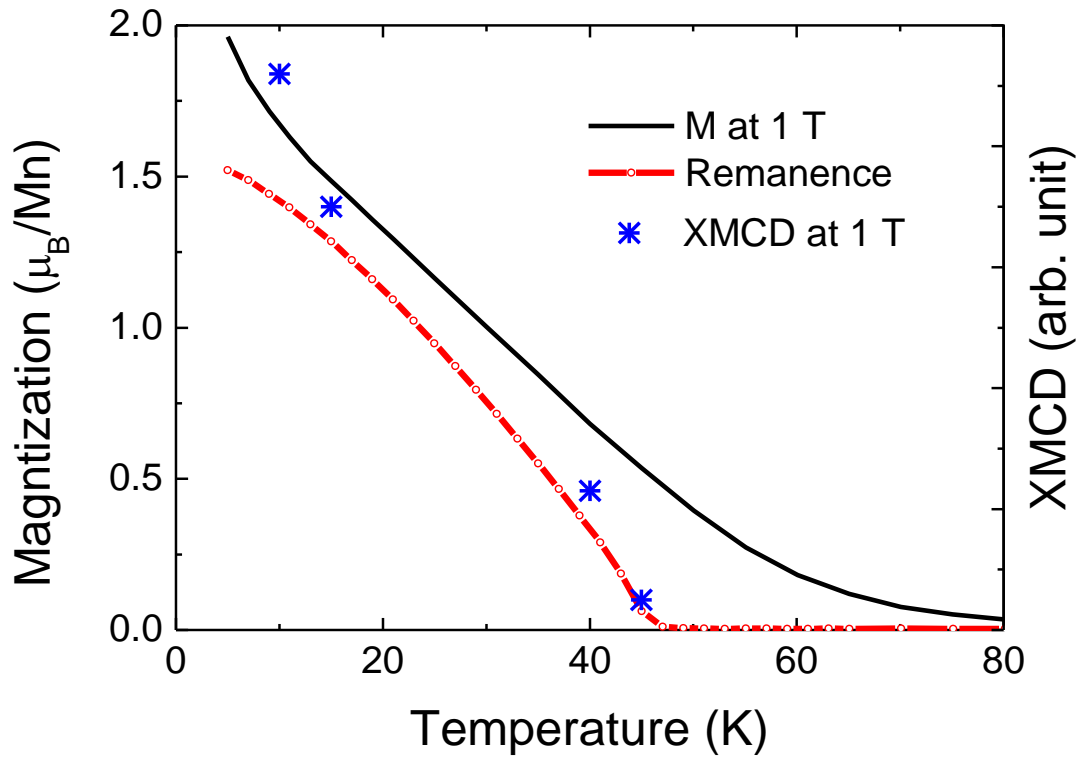

Fig. 3 Temperature dependence of magnetization measured at 0.001 T (remanence, red circle line) after saturation by 1 T and at 1 T (black solid line) and normalized XMCD signal at TEY mode measured at 1 T.

In Fig. 3, we compare the magnetization measured by SQUID magnetometry and normalized XMCD signals as a function of the measurement temperature. The magnetic remanence (*i.e.*, spontaneous magnetization) was measured at 0.001 T along the easy axis after saturation at 1 T. The spontaneous magnetization vanishes at around 46 K, while the field-induced magnetization at 1 T persists up to 80 K. Above $T_C$ (46 K), the sample is paramagnetic. The XMCD signal decreases with increasing temperature and drops to around zero at 45 K. Note that there is a small discrepancy in the temperature dependence of the magnetization and the XMCD signals. This could be due to the inhomogeneous depth profile of the Mn concentration. This fact can lead to the different temperature dependence of the magnetization between the near surface region (probed by XAS) and the bulk (measured by SQUID magnetometry). The Curie temperature (46 K) is consistent with the mean-field theory by considering a Mn concentration of ~ 4% [18].

Note that another decisive method for proving the carrier-mediated ferromagnetism is magneto-transport measurement. Below $T_C$, magnetic semiconductors reveal the anomalous Hall effect and negative magnetoresistance [6, 7]. However, the InMnAs layer cannot be measured due to the large parallel conduction from the InAs substrate.

In summary, we have synthesized ferromagnetic InMnAs, possessing the typical features of magnetic semiconductors, such as magnetic anisotropy and large XMCD. The approach combining ion implantation and pulsed laser annealing originally proposed by Scarpulla *et al.* [10, 11] presents highly versatile and industry-compatible technology. By increasing the Mn concentration and optimizing the annealing parameters, we can anticipate achieving InMnAs with even higher $T_C$.

The work was funded by the Helmholtz-Gemeinschaft Deutscher Forschungszentren (HGF-VH-NG-713).